\documentclass[twocolumn,aps,showpacs,prl,superscriptaddress]{revtex4}
\usepackage{graphicx}
\usepackage{dcolumn}
\usepackage{bm}
 \usepackage{amssymb}

\begin{document}

\title{The surface-state of the topological insulator Bi$_2$Se$_3$ revealed by cyclotron resonance.}

\author{O.E. Ayala-Valenzuela}
\affiliation{National High Magnetic Field Laboratory, Los Alamos National Laboratory, MS-E536, Los Alamos, NM 87545, USA}
\author{J.G. Analytis}
\affiliation{Geballe Laboratory for Advanced Materials and Department of Applied Physics, Stanford University, CA 94305, USA}
\affiliation{Stanford Institute for Materials and Energy Science, SLAC National Accelerator Laboratory, 2575 Sand Hill Road, Menlo Park, California 94025, USA}
\author{J.-H. Chu}
\affiliation{Geballe Laboratory for Advanced Materials and Department of Applied Physics, Stanford University, CA 94305, USA}
\affiliation{Stanford Institute for Materials and Energy Science, SLAC National Accelerator Laboratory, 2575 Sand Hill Road, Menlo Park, California 94025, USA}
\author{M.M. Altarawneh}
\affiliation{National High Magnetic Field Laboratory, Los Alamos National Laboratory, MS-E536, Los Alamos, NM 87545, USA}
\author{I. R. Fisher}
\affiliation{Geballe Laboratory for Advanced Materials and Department of Applied Physics, Stanford University, CA 94305, USA}
\affiliation{Stanford Institute for Materials and Energy Science, SLAC National Accelerator Laboratory, 2575 Sand Hill Road, Menlo Park, California 94025, USA}
\author{R.D. McDonald}\email{rmcd@lanl.gov}
\affiliation{National High Magnetic Field Laboratory, Los Alamos National Laboratory, MS-E536, Los Alamos, NM 87545, USA}

\begin{abstract}
To date transport measurements of topological insulators have been dominated by the conductivity of the bulk, leading to substantial difficulties in resolving the properties of the surface. To this end, we use high magnetic field, rf- and microwave-spectroscopy to selectively couple to the surface conductivity of Bi$_2$Se$_3$ at high frequency. In the frequency range of a few GHz we observe a crossover from quantum oscillations indicative of a small 3D Fermi surface, to cyclotron resonance indicative of a 2D surface state.
\end{abstract}


\maketitle


A massless Dirac fermion is characterized by a linear dispersion, best described in a relativistic framework and the discovery of topological insulators which support such quasiparticles has sparked substantial interest for both fundamental and technological reasons \cite{TopoReview1, TopoReview2, JoelReview}. In the last year, a large number of surface-sensitive probes have reported the existence of Dirac quasiparticles, similar to those reported in graphene \cite{mcclure_diamagnetism_1956,jiang_infrared_2007,deacon_cyclotron_2007,grapheneCR1} on the surface of single crystals of Bi$_2$Se$_3$ and related compounds \cite{Bi2Se3ARPES, Bi2Se3ARPES_a, Bi2Te3ARPES,BiSbARPES}. An experimental signature of the linear dispersion is that the cyclotron
resonance (resonant optical absorption in the presence of a magnetic field) occurs between Landau quantized energy levels $E_n$ determined by the expression \cite{mcclure_diamagnetism_1956}
\begin{equation}
  E_n = v_{\rm F}\sqrt{2(n+\gamma)\hbar e B},
\label{sqrtB}
\end{equation}
where $v_{\rm F}$ is the Fermi velocity, $n$ is the Landau level index, $\gamma$ is the Berry phase (which depends on the topology of the Fermi surface) and $B$ is the magnetic field. While recent
scanning-probe microscopy experiments have observed evidence for this unique quantization \cite{Bi2Se3STM, hanaguri_momentum-resolved_2010}, cyclotron resonance experiments \cite{STP_CR} and recent far-infrared measurements \cite{BasovCR, ButchHighMob} on the same compounds have only detected signal from the bulk. The complications associated with the presence of a bulk Fermi surface make transport and cyclotron resonance detection of the surface state significantly more difficult than in mono-layer graphene \cite{JamesCoexist}. 

In this letter, we report experiments which deconvolve bulk and surface properties of Bi$_2$Se$_3$, by employing a high frequency contactless conductivity techniques. By increasing the frequency at which the measurement is performed we reduce the skin depth and decrease our sensitivity to the bulk conduction electrons. We observe 2D cyclotron resonance phenomena, indicative of the surface state coexisting with the bulk 3D Fermi surface, as anticipated from our earlier comparison of ARPES and quantum oscillation measurements \cite{JamesCoexist}. We find that the cyclotron mass of the surface states is enhanced over that anticipated from photoemission measurements of the Fermi velocity \cite{Bi2Se3ARPES}, suggesting significant many-body renormalization of the surface Dirac fermions of this topological insulator.

\begin{figure}[ht]
\centering
\includegraphics[width=0.7\columnwidth]{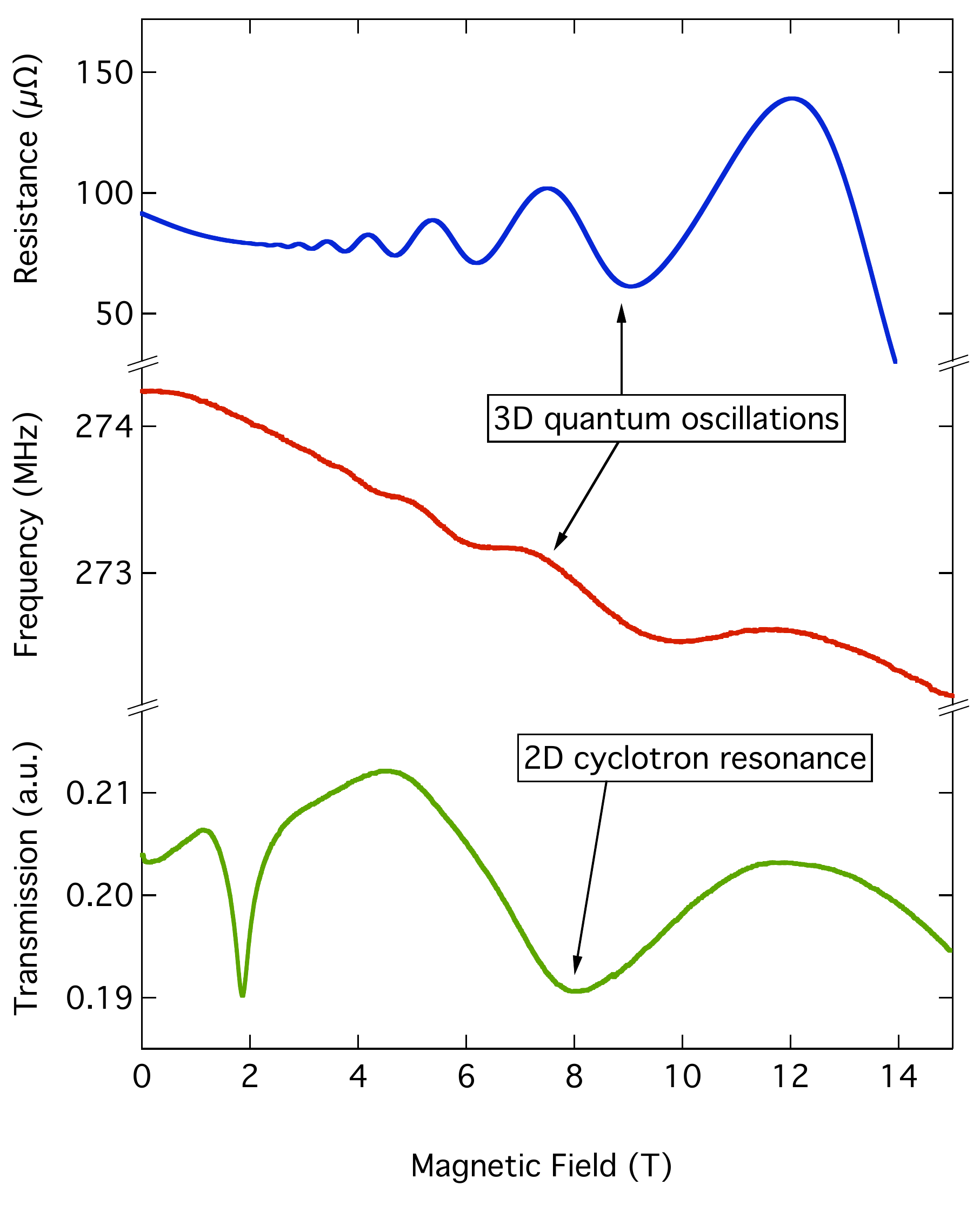}
\caption{ The magneto-conductivity of Bi$_2$Se$_3$ measured at three very different frequencies, top:  4-wire resistance measured at 312~Hz, middle: rf-skindepth (frequency shift of resonator) measured at 275~MHz and bottom: microwave cavity transmission measured at 71~GHz. All measurements were performed at a temperature of 1.5~K with the magnetic field applied parallel to the crystallographic c-axis Note that the r.f. and microwave measurements were made on the same sample, the resistance \cite{JamesCoexist} was measured on a different sample from the same growth. The microwave absorption at 2T is an isotropic cavity background. The angular dependence of the A.C. and microwave measurements are shown in Figure~\ref{Fig1b}. }
\label{Fig1}
\end{figure}

Single crystals of Bi$_2$Se$_3$ were grown by slow cooling a binary melt, conditions are described elsewhere \cite{JamesCoexist}, yielding samples with typical dimensions of $2\times1\times0.1$~mm$^3$. The trigonal c-axis is perpendicular to the cleavage plane of the crystals. Low magnetic field Hall measurements were used to determine that the bulk carrier concentrations were $2\pm0.5\times10^{17}$~cm$^{-3}$ \cite{JamesCoexist}. The resistance was measured using standard 4-wire ac transport at a frequency of  312 Hz.  

The magneto-conductivity was measured using rf- and microwave techniques. The former was achieved by inductively perturbing a resonant tank circuit \cite{MoazContactless}. The resonant frequency of the circuit is dependent upon the inductance of the sample coil, such that as the sample skin depth increases with resistance, the resonant frequency decreases. For the highest frequency circuit, 275~MHz, the skin depth is estimated from the bulk conductivity to be 30~$\mu$m, a significant fraction of the sample thickness. The conductivity at microwave frequencies was measured using a cavity perturbation technique, whereby the change in cavity transmission at resonance reflects the field induced changes in the complex conductivity of the sample. Two microwave cavities were used, one a fixed-angle multi-moded cylindrical cavity in the frequency range of 10-40~GHz \cite{NithoireaEPR}, the other a mono-moded cavity resonating at 71~GHz, that can be rotated with respect to the applied magnetic field at cryogenic temperatures \cite{ArzhanRotRes,MarijeFTR}. Both were measured using an MVNA spectrometer manufactured by AB-mm. For 71~GHz the bulk conductivity, yields a skin depth of 2~$\mu$m, only 50 times the thickness of the depletion region estimated for this carrier concentration \cite{JamesCoexist}. It should be further noted that
these estimates do not account for the possible screening effect of a high mobility surface state. The oscillating magnetic field in the coil perturbation technique is parallel to applied magnetic field and perpendicular to the (001) crystal surface so only induces screening currents in this plane. For both microwave cavity geometries the sample is located with the oscillating magnetic field in the plane of the (001) surface so induces screening currents both in and perpendicular to this plane. Magnetic fields were provided by both superconducting solenoids (up to 17~T) and by resistive Bitter magnets (up to 35~T) at the NHMFL in Tallahassee. Standard $^4$He techniques were employed to regulate temperature down to 1.5~K.

\begin{figure}[ht]
\centering
\includegraphics[width=0.7\columnwidth]{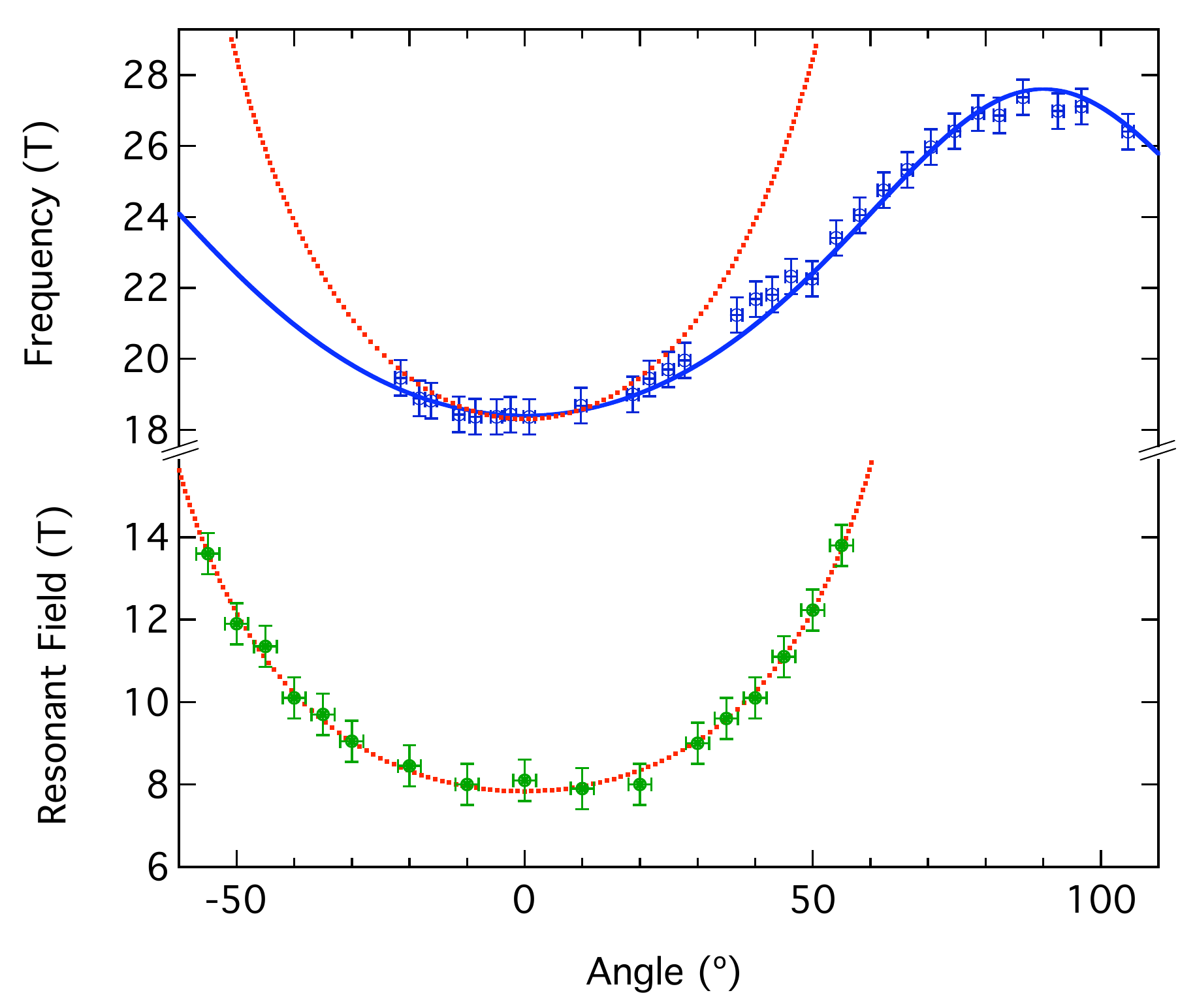}
\caption{ The angular dependence of the quantum oscillation frequency  (measured at 312~Hz) showing slightly anisotropic 3D behavior (upper points) and the angular dependence of the cyclotron resonance field  (measured at 71~GHz) showing 2D behavior (lowerpoints). Measurements correspond to the field sweeps shown in Figure~\ref{Fig1}. Both dashed lines are $B_0/cos(\theta)$, as expected for a 2D phenomenon. The solid line is $B_0/\sqrt{cos^2(\theta) + \eta^2 sin^2(\theta)}$, corresponding to a 3D Fermi surface, with an ellipticity, $\eta = 1.50$.}
\label{Fig1b}
\end{figure}

Figure~\ref{Fig1} illustrates the magneto-conductivity of Bi$_2$Se$_3$ using the three different techniques. The magneto-conductivity measured at frequencies below 275~MHz is dominated by quantum-oscillations of the bulk Fermi surface. The period of the quantum oscillations (in inverse field) is proportional to the Fermi surface cross-section, related by the Onsager equation $A_{k} = (n+\gamma)\frac{2\pi e B}{\hbar}$\cite{Onsager,Shoenberg}. From the angle dependence of the frequency of the Shubnikov-de Haas oscillations (Figure \ref{Fig1b}) we deduce that the bulk Fermi surface consists of a single closed ellipsoid, with ellipticity$<2$ \cite{JamesCoexist,Kohler}. For the same sample measured at a frequency of 71~GHz however, the signal is dominated by a broad absorption centered around 8~T. 

\begin{figure}[ht]
\centering
\includegraphics[width=.85\columnwidth]{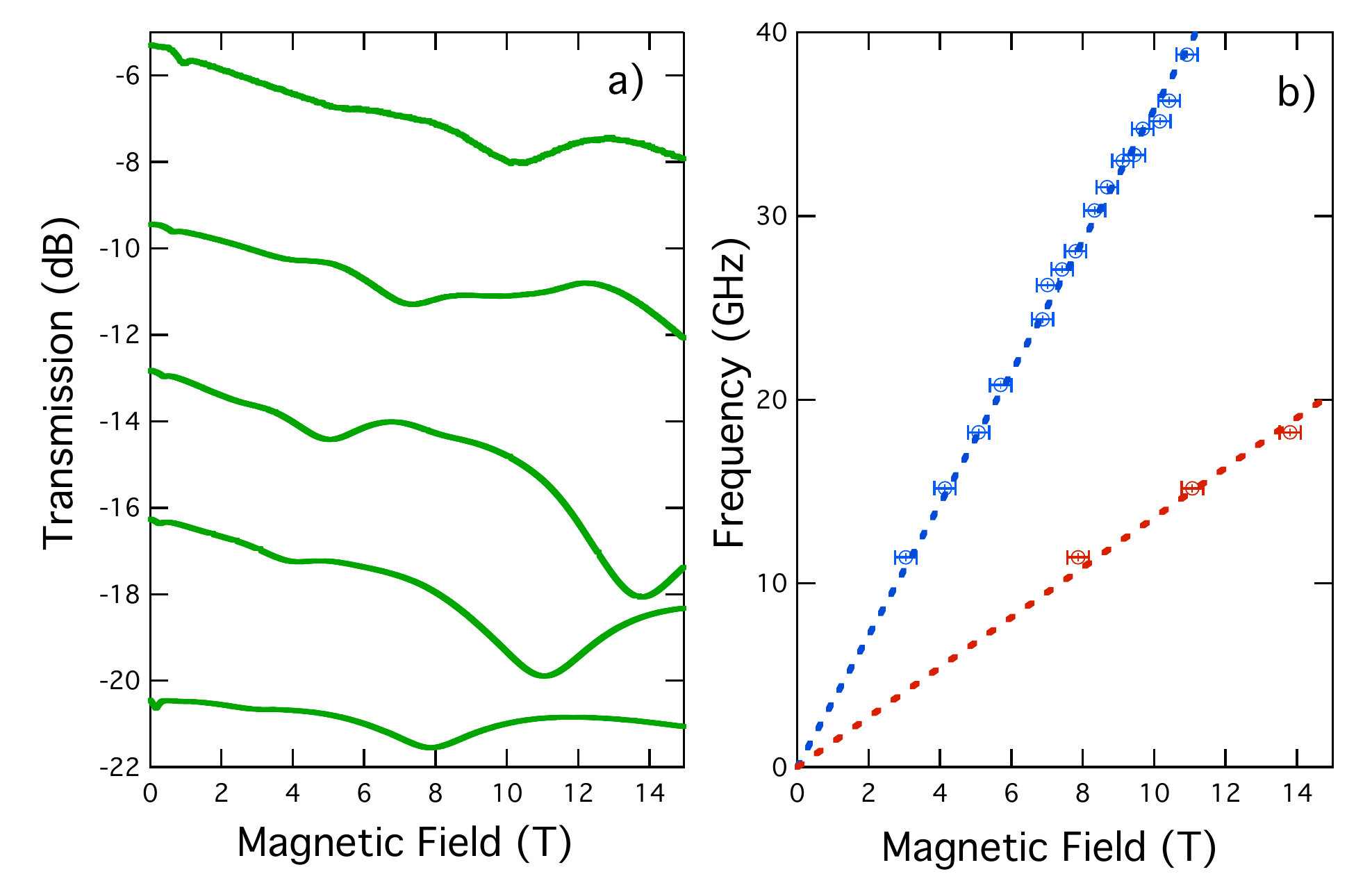}
\caption{ a) Magnetic field dependence of the cavity transmission for a selection of frequencies, top to bottom: 36.3, 27.1, 18.2, 15.1, 11.4~GHz, offset for clarity. b) The corresponding frequency vs magnetic field for the two main cyclotron resonances observed in the field range 10 to 40~GHz. It should also be noted that the two resonances observed do not have a simple harmonic relation leading us to believe that they originate from opposite sample surfaces. The factor of $\approx2$ variation in $\omega_{\rm C}$ and hence $E_{\rm F}$ between surfaces is consistent with the ARPES results from this growth reported in \cite{JamesCoexist}, see surface doping arguments below.}
\label{Fig2}
\end{figure}

\begin{figure}[ht]
\centering
\includegraphics[width=0.8\columnwidth]{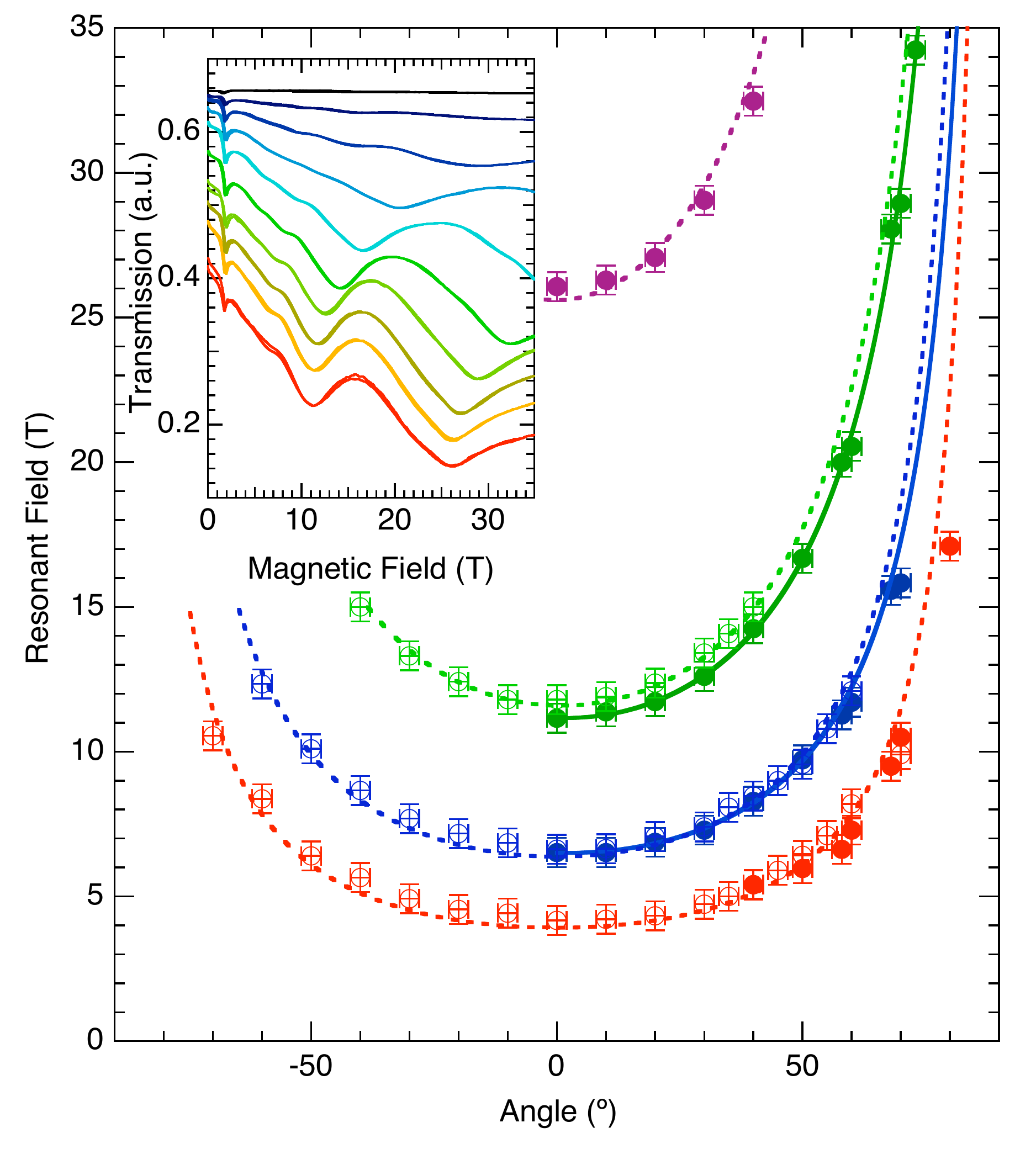}
\caption{The angular dependence of the cyclotron resonance field and harmonics measured at 71~GHz. Open symbols are data points measured in Los Alamos, solid symbols are a subsequent measurement on the same sample in Tallahassee, note the systematic shift to lower field with time. 
The lines are fits to $B_0/cos(\theta)$ indicating the 2D nature of the resonance. The inset shows examples of transmission data offset for clarity, top to bottom is 90$^\circ$ to 0$^\circ$ in 10$^\circ$ steps. Note that the reduced amplitude at high angles is due to the angular dependence of the cavity coupling. }
\label{Fig3}
\end{figure}

For a different sample from the same growth, Figure~\ref{Fig2}a) plots the transmission measured at different microwave frequencies, indicating that this is a resonant phenomenon with linear frequency field scaling. The wide field range over which data from the sample in Figure~\ref{Fig3} was collected leads us to identify the resonances as a single harmonic series, the weaker, lower field resonances occurring at magnetic fields close to $\frac{1}{2}$, $\frac{1}{3}$ and  $\frac{1}{4}$ that of the fundamental. The frequency field scaling of the fundamental cyclotron resonance condition is 2.7~GHzT$^{-1}$, providing compelling evidence that this is not cyclotron resonance originating from quasi-particles with an effective mass the same as the conduction band, $m^\star = 0.12~m_e$, \footnote{previous studies indicate that the conduction band mass is only very weakly dependent upon band filling \cite{Kohler}} which would lead to a resonance condition corresponding to 233~GHzT$^{-1}$, around 2 orders of magnitude greater than our observation. Neither can this be a cyclotron resonance harmonic $\omega = l \omega_{\rm c}$, from a band with a mass of $0.12~m_e$, as at 71 ~GHz one would expect subsequent harmonics to be less than 1~T apart in field. This resonance therefore does not originate from the bulk band structure, and the remainder of this study is dedicated to the investigation of these high-frequency features.

Figures~\ref{Fig1b} and \ref{Fig3} show the angle dependence of the high-frequency resonance for two different samples, where the angle $\theta$ is defined as that between the normal to the cleavage plane (the [001] crystal axis) and the field direction. The resonance field depends only on the component of the field which is perpendicular to the cleavage plane and scales as $1/cos(\theta)$. This is an unambiguous signature of a 2D state. As this property only emerges at our highest frequencies where the skin depth is smallest, it is natural to attribute the origin of this signal to the 2D Dirac fermions,
known to exist on the surface of  the material from photoemission experiments \cite{Bi2Se3ARPES_a,JamesCoexist}.

As we show below the observed frequency scaling, that is linear in $B$ rather than $\sqrt{B}$, as perhaps expected from Equation \ref{sqrtB}, is consistent with the Fermi energy relative to the Dirac point being large in comparison to the temperature and measurement frequency. To model this behavior, and confirm that the linear regime is expected for these samples, the absorption, $\alpha$, is approximated as the sum of Lorentzian lines centered upon the frequency corresponding to adjacent Landau level separation. The intensity for each transition is weighted by the probability of occupation of the initial and final states, thus ensuring only transitions from occupied to unoccupied states contribute.
\begin{equation}
\alpha(B) = \sum_n \left[ \frac{\tau}{(\hbar \omega - E_n - E_{n+1})^2 +\tau^2}\right] F(E_n) [1-F(E_{n+1})],
\label{cycampl}
\end{equation}
where $F(E)$ is the Fermi-Dirac distribution function and $E_n$ is given by equation~\ref{sqrtB}. The magnetic field corresponding to the fundamental cyclotron resonance condition was calculated as the maximum in absorption for a given frequency. The calculation was repeated for a range of frequencies and for higher harmonics corresponding to $\Delta n \neq \pm1$.  For absorption line widths, $\tau$, less than the experimental temperature the result of the simulation was found independent of the value used and for a large Fermi energy, the resonance condition is independent of the value for the additional Berry phase parameter, $\gamma$.  Unlike the case for massive quasi-particles where the gradient of the resonant frequency/magnetic field condition is determined solely by the effective mass, in this case it is determined by both the Fermi-velocity and -energy,
\begin{equation}
\omega_{\rm c} = \frac{eB v_{\rm F}^2}{E_{\rm F}}
\label{linB}
\end{equation}
The simulation yields the same result as evaluating the cyclotron mass \cite{Shoenberg}
\begin{equation}
m_{\rm c}(E_{\rm F}) = \frac{\hbar^{2}}{2 \pi}  \left. \frac{dA_{k}}{dE}\right |_{E_{\rm F}}
\label{cycmass}
\end{equation}
at a constant energy ($E_{\rm F}$) as opposed to the non-constant energy limit that yields the $\sqrt{B}$ field dependence of the cyclotron frequency for band fillings close to the Dirac point. This case is satisfied in  graphene\cite{jiang_infrared_2007,deacon_cyclotron_2007,grapheneCR1}, but in the present case with high filling, we expect the linear $B$ dependence of Equation \ref{linB}.

ARPES \cite{JamesCoexist,Bi2Se3ARPES,Bi2Se3ARPES_a} measures the velocity associated with the surface state to be $4.2\times10^5$ms$^{-1}$ and puts the Fermi energy close to the bulk conduction band edge, {\it i.e.} approximately 200~meV above the Dirac point. However, the Fermi surface of the Dirac fermion is set by surface charge accumulation which depends sensitively on the atmospheric exposure of the sample \cite{JamesCoexist}. This leads to sample and exposure dependence of the 2D cyclotron resonance which depends on the Fermi energy (Equation \ref{linB}), as shown in Figure \ref{Fig3} and by the different cyclotron resonance conditions observed for the 3 different samples in Figures~\ref{Fig1}, \ref{Fig2} and \ref{Fig3}. In contrast, this behavior would not be expected from a bulk cyclotron resonance, which depends only on the effective mass.

Using $E_{\rm F} = 200$~meV requires a Fermi velocity of the Dirac cone nearly 7 times smaller than that measured by ARPES, suggesting substantial many-body renormalization of the topological surface state Dirac fermions. Recent far-infrared reflection measurements \cite{BasovCR} indicate strong electron-phonon coupling, consistent with the large dielectric constant \cite{ButchHighMob}. The authors also report coupling between these optic phonons in the 10~meV range and magnetic field \cite{BasovCR}. Although electron-electron interactions do not usually renormalize the cyclotron mass, it has been reported \cite{MarijeFTR} that the thermodynamic effective mass, that is probed by the temperature dependance of quantum oscillations \cite{MarijeFTR}, can be exceeded by the cyclotron mass. The observation of an electron-electron enhanced cyclotron mass is also in agreement with Hubbard-model calculations for interacting systems \cite{KankiYamada}. Recent predictions for the surface state collective mode in topological insulators \cite{HelicalCollective} and in particular Bi$_2$Se$_3$ indicate that not only does the spin wave couple to sound propagation but that Coulomb interactions can couple the spin wave and surface plasmon modes. For the region of their dispersion where they coexist, it is reasonable to speculate whether self-energy corrections due to excitation of this spin-plasmon will enhance the observed cyclotron mass over that of the bare Dirac dispersion. These measurements hence place important constraints upon the magnitude of these effects. 

In conclusion, by probing the conductivity at reduced skin depths, we have observed a 2D cyclotron resonance from a material whose bulk Fermi-surface is 3D. The frequency-magnetic field scaling of this resonance is inconsistent with the bulk effective mass, but more consistent with the dispersion and band filling of a Dirac-like surface state as observed by ARPES \cite{JamesCoexist}, with substantial many-body renormalization.

We would like to thank Steve Hill and Saiti Data for technical assistance with the Tallahassee MVNA. RMcD acknowledges support from the U.S. DOE, Office of Basic Energy Sciences `Science in 100 T' program. Work at Stanford was supported by the U.S. DOE, Office of Basic Energy Sciences under contract DE-AC02-76SF00515.

\begin{thebibliography}{1}

\bibitem{TopoReview1} M.Z.~Hasan and C.L.~Kane, arXiv:{\bf 1002.3895} (2010)

\bibitem{TopoReview2} L.~Fu and C.L.~Kane, Phys. Rev. B, {\bf 76}, 045302 (2007)

\bibitem{JoelReview} J. E. Moore, Nature {\bf 464}, 194 (2010). 

\bibitem{mcclure_diamagnetism_1956} J. W. {McClure}, Physical Review {\bf 104}, 666 (1956)

\bibitem{jiang_infrared_2007} Z.  Jiang and E. A. Henriksen and L. C. Tung and {Y.-J.}  Wang and M. E. Schwartz and M. Y. Han and P.  Kim and H. L. Stormer, Physical Review Letters {\bf 98}, 197403 (2007)

\bibitem{deacon_cyclotron_2007} R. S. Deacon and {K.-C.}  Chuang and R. J. Nicholas and K. S. Novoselov and A. K. Geim, Physical Review B {\bf 76}, 081406 (2007)

\bibitem{grapheneCR1} K.S.~Novoselov, A.K.Geim, S.V.~Morozov, D.~Jiang, M.I.~Katsnelson, I.V.~Grigorieva, S.V. Dubonos and A.A.~Firsov, Nature {\bf 438} 197 (2005)

\bibitem{Bi2Se3ARPES} D.~Hsieh, Y.~Xia, D.~Qian, L.~Wray, J.H.~Dil, F.~Meier, J.~Osterwalder, L.~Patthey, J.G.~Checkelsky, N.P.~Ong, A.V.~Fedorov, H.~Lin, A~Bansil, D.~Grauer, Y.S.~Hor, R.J.~Cava and M.Z.~Hasan, Nature {\bf 460}, 1101 (2009) 

\bibitem{Bi2Se3ARPES_a} Y. Xia and D. Qian and D. Hsieh and L. Wray and A. Pal and H. Lin and A. Bansil and D. Grauer and Y. S. Hor and R. J. Cava and M. Z. Hasan, Nature Physics {\bf 5}, 398 (2009)

\bibitem{Bi2Te3ARPES} Y.L.~Chen, J.G.~Analytis, J.-H.~Chu, Z.K.~Liu, S.-K.~Mo, X.L.~Qi, H.J.~Zhang, D.H.~Lu, X.~Dai, Z.~Fang, S.C.~Zhang, I.R.~Fisher, Z.~Hussain, and Z.-X.~Shen, Science {\bf 325}, 178 (2009)

\bibitem{BiSbARPES} D.~Hsieh, Y.~Xia, L.~Wray, D.~Qian, A.~Pal, J.H.~Dil, J.~Osterwalder, F.~Meier, G.~Bihlmayer, C.L.~Kane, Y.S.~Hor, R.J.~Cava, and M.Z.~Hasan, Science {\bf 323}, 919 (2009)

\bibitem{Bi2Se3STM} P.~Cheng, C.~Song, T.~Zhang, Y.~Zhang, Y.~Wang, J-F.~Jia, J.~Wang, Y.~Wang, B-F.~Zhu, X.~Chen, X.M.K.~He, L.~Wang, X.~Dai, Z.~Fang, X.C.~Xie, X-L.~Qi, C-X.~Liu, S-C.~Zhang and Q-K.~Xue, arXiv:{\bf 1001.3220} (2010)


\bibitem{hanaguri_momentum-resolved_2010} T. Hanaguri and K. Igarashi and M. Kawamura and H. Takagi and T. Sasagawa, arXiv:{\bf 1003.0100}

\bibitem{STP_CR} V.A.~Kulbachinskii, N.~Miura, H.~Nakagawa, H.~Arimoto, T.~Ikaida, P. Lostak and C. Drasar, Phys. Rev. B. {\bf 59} 15733 (1999)

\bibitem{BasovCR} A.D.~LaForge, A.~Frenzel, B.C.~Pursley, T.~Lin, X.~Liu, J.~Shi, and D.N.~Basov, arXiv:{\bf 0912.2769} (2009)

\bibitem{ButchHighMob} N.P.~Butch, K.~Kirshenbaum, P.~Syers, A.B.~Sushkov, G.S.~jenkins, H.D.~Drew and J.~Paglione, arXiv:{\bf 1003.2382} (2010)

\bibitem{JamesCoexist} J.G.~Analytis, J-H.~Chu, Y.~Chen, F.~Corredor, R.D.~McDonald, Z.X.~Shen and I.R.~Fisher, arXiv:{\bf 1001.4050} (2010)

\bibitem{MoazContactless} M.M.~Altarawneh, C.H.~Mielke, and J.S.~Brooks, Rev. Sci. Inst. {\bf 80} 066104 (2009)

\bibitem{NithoireaEPR} S.~Cox, R.D.~McDonald, M.~Armanious, P.~Sengupta, and A.~Paduan-Filho, Phys. Rev. Lett. {\bf 101}, 087602 (2008)

\bibitem{ArzhanRotRes} A.~Ardavan, R.S.~Edwards, E.~Rzepniewski, J.~Singleton and S.J.~Blundell, Physica B, {\bf 294-295}, 379 (2001) 

\bibitem{MarijeFTR} J.M.~Schrama, J.~Singleton, R.S.~Edwards, A.~Ardavan, E.~Rzepniewski, R.~Harris, P.~Goy, M.~Gross, J.~Schlueter, M.~Kurmoo and P.~Day,  J. Phys.: Cond. Matt {\bf 13}  2235 (2001)

\bibitem{Onsager} L.~Onsager, Philos. Mag. {\bf 43}, 1006 (1952)

\bibitem{Shoenberg} D.~Shoenberg, Magnetic Oscillations in Metals (Cambridge University Press, Cambridge, England, 1984)


\bibitem{KankiYamada} K.~ Kanki and K.~ Yamada, J.~Phys.~Soc.~Jap. {\bf 66} 1103 (1997)

\bibitem{HelicalCollective}S.~Raghu, S.B.~Chung, X-L.~Qi, S-C.~Zhang arXiv:{\bf 0909. 2477} (2010)

\bibitem{Kohler}H. Kohler, Physica Status Solidi (b) {\bf 58}, 91 (1973)

\end{thebibliography}

\end{document}